\documentclass[journal,10pt]{IEEEtran}

\usepackage{CJK}
\usepackage{cite}
\usepackage[justification=centering]{caption}
\usepackage{graphicx}
\usepackage{epstopdf}
\usepackage{subfigure}
\usepackage[cmex10]{amsmath}
\usepackage[amsmath,thmmarks]{ntheorem}
\usepackage{amssymb}
\usepackage{ntheorem}
\usepackage{mathabx}
\usepackage{algorithm,float}
\usepackage{algorithmic}
\usepackage{lipsum}
\usepackage{cuted}
\usepackage{booktabs}
\usepackage{threeparttable}
\usepackage{multicol}
\usepackage{multirow}
\usepackage{xcolor}
\usepackage{bm}
\usepackage{mathabx}
\usepackage{stmaryrd}
\usepackage{upgreek}
\usepackage{url}
\usepackage{wasysym}
\usepackage[labelsep=none]{caption}
\usepackage{lettrine}

\begin{document}


\title{Time-Frequency Analysis based Deep Interference Classification for Frequency Hopping System}
\author{Changzhi Xu, Jingya Ren, Wanxin Yu, Yi Jin, Zhenxin Cao, Xiaogang Wu and Weiheng Jiang
%
\thanks{
C. Xu is with State Key Laboratory of Millimeter Wave, Southeast University, Nanjing, China, and he is also with Xi'an Space Radio Technology Institute, Xi'an, China (e-mail: sandy\_xu@126.com).

J. Ren, W. Yu, X. Wu and W. Jiang are with the School of Microelectronics and Communication Engineering, Chongqing University, Chongqing, China (e-mail: 202112131071t@cqu.edu.cn, xiaogangwu@cqu.edu.cn, wxyuwan@cqu.edu.cn, whjiang@cqu.edu.cn).

Y. Jin is with Xi'an Space Radio Technology Institute, Xi'an, China (e-mail: john.0216@163.com).

Z. Cao is with State Key Laboratory of Millimeter Waves Southeast University, Nanjing, China (caozx@seu.edu.cn).
}
}


\maketitle

\begin{abstract}
It is known that, interference classification plays an important role in protecting the authorized communication system and avoiding its performance degradation in the hostile environment. In this paper, the interference classification problem for the frequency hopping communication system is discussed. Considering the possibility of presence multiple interferences in the frequency hopping system, in order to fully extract effective features of the interferences from the received signals, the linear and bilinear transform based composite time-frequency analysis method is adopted. Then the time-frequency spectrograms obtained from the time-frequency analysis are constructed as matching pairs and input to the deep neural network for classification. In particular, the Siamese neural network is used as the classifier, where the paired spectrograms are input into the two sub-networks of the deep networks, and these two sub-networks extract the features of the paired spectrograms for interference type classification. The simulation results confirm that the proposed algorithm can obtain higher classification accuracy than both traditional single time-frequency representation based approach and the AlexNet transfer learning or convolutional neural network based methods.
\end{abstract}

\begin{IEEEkeywords}
Interference Classification, Time-frequency Analysis, Siamese Neural Network, Frequency Hopping
\end{IEEEkeywords}
\IEEEpeerreviewmaketitle

\section{Introduction}
Anti-jamming plays an important role in protecting wireless based communication system, especially in the hostile environment \cite{1}, and interference classification is the prerequisite for realizing the anti-jamming. That is, only by knowing the type of interference that the wireless system is suffering, then it is possible to develop effective method for interference suppression or elimination. Till now, though lots of works have been done for the interference classification problem and many algorithms have been proposed for various wireless systems \cite{2,3,4,5,6,7,8,9,10,11,12,13,14,15,16,17,18,19,20}. However, currently, the emergence of many new interference waveforms seriously degrade the performance of these algorithms \cite{15,16,17,18,19,20}. At the same time, with the development of deep learning these years \cite{21}, the capability of extracting effective features from samples has been enhanced. Therefore, with deep learning, how to design interference classification algorithm with higher accuracy is an important research topic, especially in the complicated interference environment, which is the focus of this paper.


By considering the systems' coexisting over the ISM bands, interference classification and avoiding is an important issue for these systems which has been well discussed yet. Specifically, in \cite{3}, the WLAN interference classification under factory environments was discussed, where scalogram time frequency images were computed from the collected received signal strength (RSS) data and then a convolutional neural network (CNN) was trained to recognize the spectral features and enable the interference classification. In \cite{4}, the authors presented a semi-supervised deep learning (DL) based wireless interference identification (WII) algorithm which combined temporal ensembling technique with CNN network to exploit unlabeled data to improve the performance. In \cite{5}, a deep neural network based interference-classification method was proposed, in which both the power-spectral density (PSD) and the cyclic spectrum of the received signal were treated as input features to the network. The computer experiments showed that the accuracy with the received signal PSD outperforms that with its cyclic spectrum. Motivated by the increasing need for on-device interference detection and identification (IDI) for wireless coexistence, \cite{6} developed a lightweight and efficient method targeting interference identification already at the level of single interference bursts which exploited real-time extraction of envelope and model-aided spectral features, specifically designed considering the physical properties of the signals captured with commercial off-the-shelf (COTS) hardware. \cite{7} studied the problem of interference source identification, through the lens of recognizing one of 15 different channels under which 3 different wireless technologies, i.e., Bluetooth, Zigbee, and WiFi. A few works have demonstrated the effectiveness of deep learning in classifications, such as CNN, ResNet, CLDNN, and LSTM \cite{8,9,10,11}. In \cite{8}, a real-time external interference source classification method for an 802.15.4-based wireless sensor network using convolutional neural network was proposed, which used RSSI sampling for collecting training and test data in an office environment,
experimental results confirmed that the proposed framework can classify the major interference types with high accuracy. Similarly, the authors of \cite{9,10} proposed a wireless interference identification (WII) approach based upon a deep convolutional neural network (CNN) which classified multiple IEEE 802.15.1, IEEE 802.11 b/g and IEEE 802.15.4 interfering signals in the presence of a utilized signal, with a classification accuracy of approximately 90 \% at least.
In \cite{11}, an unsupervised learning method and the unknown interference classifier were proposed based on the self-organizing map (SOM) neural network, the simulation results demonstrated that, when the SNR reached 5 dB, the accuracy of unknown interference classification exceeded 94\%. More studies about the interference identification problems for WLAN and cellular systems can be found in \cite{12,13,14,1Contour} for the interference identification in the WiFi system, inter-femtocell network and the C-RAN networks, respectively.

Besides the studies for traditional commercial networks, the interference classification for private or special wireless networks also has drawn wide attentions, especially for the applications in the hostile or adversarial environments, i.e., the frequency hopping or the direct sequence spreading spectrum (DSSS) based system. Specifically, in \cite{15}, for the time-slotted channel hopping (TSCH)-based industrial wireless sensor and actuator networks (IWSANs), a centralized interference classifier based on support vector machines (SVMs) was introduced. In \cite{16}, an iterative anti-interference method based on interference power cognition was proposed, where the interference was broadband digital modulation signal and desired signal was direct sequence spread spectrum(SS) signal. In \cite{17}, a blind user identification detection (UID) and interference identification scheme based on linear prediction algorithm for asynchronous direct-sequence code-division multiple-access systems over multipath fading channels was solved. Similarly, the work in \cite{18} extracted 3dB bandwidth, time domain peak-to-average ratio and other features from the time-frequency domain information of the interference signal, and used decision tree and deep network classifiers to complete the classification of the interference signal. \cite{19} proposed an interference recognition scheme based on a self-organizing map neural network, which improved the classification accuracy of known interference by 3.44\%, and reached an accuracy of 94\% when the interference was unknown at the SNR of 5dB. Different from the aforementioned methods while the traditional statistical characteristics are calculated as classification features, \cite{20} used linear time-frequency analysis to extract time-frequency features of the mixed interference signals and then the transfer learning based classifier was designed, it has higher accuracy but with a problem of lower time-frequency resolution during signal interleaving, and also the demand for training samples is higher.

We observe that there are still many shortcomings for the above-mentioned existing studies, especially for the frequency hopping wireless communication systems, which motivates the work of this paper. In practice and for the frequency hopping system, there may be multiple interferences presented in the system, i.e., in hostile environment with adversarial interference waveform. Under this condition, the existing algorithms which only calculate traditional higher-order statistics or single time-frequency characteristics may become invalid.
Therefore, in order to improve the classification accuracy over complexity environments, especially for the frequency hopping system with multiple or composite interferences, we try to explore more effective methods for feature extracting and classification. In this paper, the interference classification problem for the frequency hopping communication systems is discussed and then a composite time-frequency analysis and deep learning based algorithm is proposed. Specifically, the contributions of this paper are summarized as follows.
\begin{itemize}
  \item Considering the presence of multiple and compound interference in the frequency hopping system, in order to fully extract effective features of the interferences from the received signals to perform high precision interference classification, a composite time-frequency analysis method based on both the linear transformation and bilinear transformation is proposed.
  \item In addition, in order to realize high-precision interference classification with small samples, the Siamese neural network is adopted as the classifier, where the paired spectrograms are input into two sub-networks of the Siamese neural network, and these two sub-networks extract the features of the paired spectrograms. Then the Siamese neural network is trained and tested for interference type classification via calculating the gap between the generated features.
\end{itemize}

The rest of this paper is organized as follows. We briefly introduce the process of the proposed interference classification algorithm in Section II. The details of the preprocessing steps for the interference classification algorithm, i.e., the composite time-frequency analysis, normalization, binarization, cropping and resizing are presented in Section III. We illustrate the architecture of the used Siamese neural network and its pre-training process in Section IV. The performance of the proposed algorithm is evaluated and analyzed in Section V and then we make conclusions at last.

\section{Flow Chart of Interference Classification}
A typical frequency hopping communication system includes a frequency hopping transmitter, wireless channel and a frequency hopping receiver, where the frequency hopping transmitter radiates the signal through the antenna after preprocessing and frequency hopping modulation. The frequency hopping signal transmitted over the wireless channel would be affected by noise and interference. Therefore, the signal received by frequency hopping receiver is composed of frequency hopping modulated signal, interference and noise as [2]-[4], [18]

\begin{equation}
r\left(t\right)={s_1}\left(t\right)+\sum\limits_{j=1}^N{{J_j}\left( t\right)}+n\left(t\right),
\label{eq:1}
\end{equation}
where ${s_1}(t)$ is the received desired frequency hopping signal,
${J_j}(t),i=1,\ldots,N$ denote the potential interference signals received at the receiver, which including fixed interference, periodic pulse interference, comb spectrum interference, periodic sweep interference and the mixed interferences. Specifically, the forms of these interference signals are given in the Appendix, and $n(t)$ denotes the noise signal. 

\begin{figure*}[!ht]
\centerline{\includegraphics[scale=0.77]{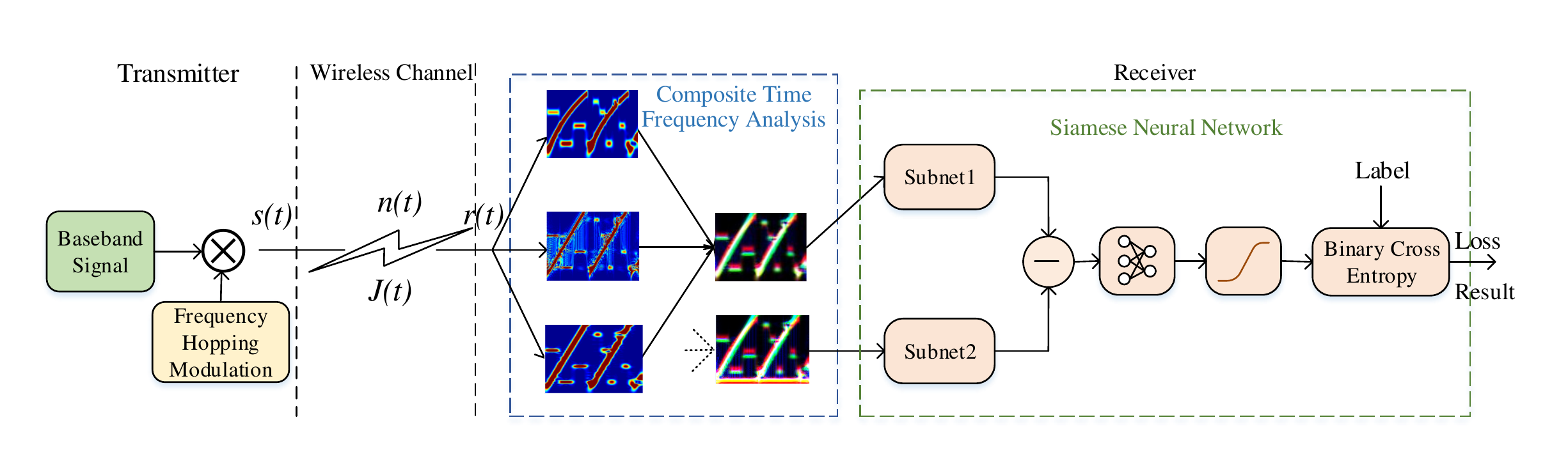}}
\captionsetup{font={footnotesize}}
\caption{$~$Flow chart of interference classification.}
\label{Fig.1.}
\end{figure*}

In general, with the receiving signal (1), the frequency hopping receiver directly performs decoding after demodulation. However, if interference presents in the system, the receiver will face severe interferences and direct signal demodulation may experience worse decoding performance \cite{18}. Therefore, in order to protect the system and avoid performance degradation caused by potential interference, we need to perform interference classification and then interference suppression. While interference classification clarify the type of interference and obtain the parameters of the interference signals. As mentioned earlier, in this paper, we focus on identifying the type of interference signals presented in the system. The proposed classification process is shown in Fig. \ref{Fig.1.}. At first, we implement a composite time-frequency analysis to obtain the spectrograms of the received signals. Based on that, we then construct the matching pairs of the obtained spectrograms, and the paired spectrograms are input into two sub-networks of the Siamese neural network. These two sub-networks extract the features of the paired spectrograms. Finally, the Siamese neural network is trained and tested for interference classification by calculating the gap between the generated features. In the sequel, the details of the composite time-frequency analysis and Siamese neural network based interference classification method are presented.

\section{Composite Time-frequency Analysis}
As mentioned earlier, in order to implement interference classification, we should extract effective features from the interference signal. However, as we know that, if the interference signals are compounded with different types and also they are coincided with the frequency hopping signal, interference classification based on signal statistical characteristics such as instantaneous or high-order statistics will be greatly affected. This comes from the fact that, with compound interference signals, various time-domain, frequency-domain and instantaneous features of these interference signals and also the frequency hopping signals are affected or even canceled by each other \cite{3Efficient}. Thus it is difficult to effectively extract the typical time or frequency-domain features for these signals. To handle this issue, we propose a multi-channel composite time-frequency based method to achieve the feature extraction for the compound interference signals to assist interference classification. The core idea is that, both linear and bilinear time-frequency analysis methods are used to calculate the spectrograms of the received signals, and these spectrograms are used as features for classification. This is similar to the colored image recognition with `RGB' multiple channels, three carefully chosen time-frequency analysis methods are adopted to formulate the `RGB' channels for the interference classification, which takes the advantages of the complementarity of the capabilities of different time-frequency analysis tools. In this section, the details of the proposed multi-channel composite time-frequency method are presented.

\subsection{Linear and Bilinear Time-Frequency Transforms}
In this paper, both the linear and bilinear transform time-frequency methods are used to extract the features of the received signals. For the linear time-frequency transform, though there is no cross term, due to the existence of window functions, the resolution of time-frequency transform will be limited. Taking the short-time Fourier transform (STFT) as an example, the window function of STFT will affect both time resolution and frequency resolution, therefore, it is difficult to choose an optimal window function and set the appropriate parameters. Compared with STFT, wavelet transform can more effectively focus on the instantaneous structure of the signal \cite{Cohen1966Generalized}. In addition, compared with the linear time-frequency transform, though the Cohen-like bilinear time-frequency transform introduces the cross terms, different time-frequency analysis effects can be obtained with different kernel functions. Therefore, in order to sufficiently extract enough time-frequency features, both the linear and the Cohen-like bilinear time-frequency transforms are adopted, i.e., we choose wavelet transform, MHD and BJD bilinear time-frequency transforms to obtain the spectrograms of the composite interference signals as follows, separately.

\subsubsection{Wavelet Transform} Wavelets are wave-like transients that can be interpreted as sinusoids for short duration. Decompositions of a signal on the basis functions are called wavelet transforms (WTs). For the WTs, the width of the wavelet basis changes with frequency, i.e., if the frequency of the basis becomes larger, the time window width will automatically be narrowed to improve the resolution. Its principle is similar to a zoom camera \cite{1983ZhPhy..43..172H}. Define the wavelet transform of a signal $s(t)$ as \cite{8742635}
\begin{equation}
CW{T_s}\left( {a,b} \right) = \frac{1}{{\sqrt a }}\int_{ - \infty }^{ + \infty } {s\left( t \right)\varphi \left( {\frac{{t - b}}{a}} \right)dt},
\label{eq:2}
\end{equation}
where $a$ and $b$ are the scale (dilation) and translation parameters, respectively, and the function $\varphi(t)$ is the basis function called the mother wavelet, and it is similar to the window function of the STFT. $\varphi({\frac{{t - b}}{a}})$ is obtained by $\varphi(t)$ after translation and expansion transformation, and the basis function $\varphi(t)$ satisfies the following condition.

\begin{equation}
\int_{ - \infty }^{ + \infty } {{{\left| {\varphi \left( t \right)} \right|}^2}dt}  < \infty.
\label{eq:3}
\end{equation}

In this paper, for simplification, the commonly Morlet wavelet basis function is used which has the form

\begin{equation}
\varphi \left( t \right) = \frac{1}{{\sqrt[4]{\pi }}}{e^{j{w_0}t - \frac{{{t^2}}}{2}}}.
\label{eq:4}
\end{equation}

\subsubsection{Margenau-Hill Distribution (MHD) Bilinear Transform} It is a time-frequency analysis method with many excellent features. Specifically, MHD has true marginality, weak supporting and better time-frequency aggregation \cite{7105431}. MHD based bilinear transform can be characterized as
\begin{equation}
\begin{aligned}
MH{D_s}\left( {t,f} \right) = \iiint_{-\infty}^{+\infty}&{s\left( {t + {\raise0.7ex\hbox{$\tau $} \!\mathord{\left/
 {\vphantom {\tau  2}}\right.\kern-\nulldelimiterspace}
\!\lower0.7ex\hbox{$2$}}} \right)} {s^ * }\left( {t - {\raise0.7ex\hbox{$\tau $} \!\mathord{\left/
 {\vphantom {\tau  2}}\right.\kern-\nulldelimiterspace}
\!\lower0.7ex\hbox{$2$}}} \right)\cos \left( {{{\eta \tau } \mathord{\left/
 {\vphantom {{\eta \tau } 2}} \right.
 \kern-\nulldelimiterspace} 2}} \right)\\
&\cdot{e^{ - j2\pi \left( {\eta t + f\tau  - \eta u} \right)}}d\tau.
\end{aligned}
\label{eq:5}
\end{equation}
From the above, we can note that the kernel function of MHD is $\phi(\eta,\tau)=\cos(\eta\tau/2)$ and it is more complexity than that for Wigner-Ville distribution (WVD). Similar to pseudo-Wigner-Ville distribution (PWVD), if the time domain window function $h(\tau)$ is added to the time domain variable $\tau$, the cross term of MHD can be suppressed to a certain extent and which forms a pseudo Margenau-Hill distribution (PMHD).

\subsubsection{Born-Jordan distribution (BJD) Bilinear Transform} The kernel function of BJD is $\phi({\eta,\tau})=\sin({\pi\tau\eta})\slash{\pi \tau \eta}$. Compared with other Cohen-like time-frequency analysis methods, BJD has higher time-frequency resolution. Its expression is

\begin{equation}
\begin{aligned}
BJ{D_s}\left( {t,f} \right) = \frac{1}{{2a}}\int_{ - \infty }^{ + \infty } \frac{1}{{\left| \tau  \right|}}\int_{t - a\left| \tau  \right|}^{t + a\left| \tau  \right|} &{s\left( {u + {\raise0.7ex\hbox{$\tau $} \!\mathord{\left/
 {\vphantom {\tau  2}}\right.\kern-\nulldelimiterspace}
\!\lower0.7ex\hbox{$2$}}} \right)} {s^ * }\left( {u - {\raise0.7ex\hbox{$\tau $} \!\mathord{\left/
 {\vphantom {\tau  2}}\right.\kern-\nulldelimiterspace}
\!\lower0.7ex\hbox{$2$}}} \right)\\
&\cdot{e^{ - j2\pi f\tau }}dud\tau,
\end{aligned}
\label{eq:6}
\end{equation}
where $a$ is constant.

\begin{figure}[h]
\centerline{\includegraphics[scale=0.57]{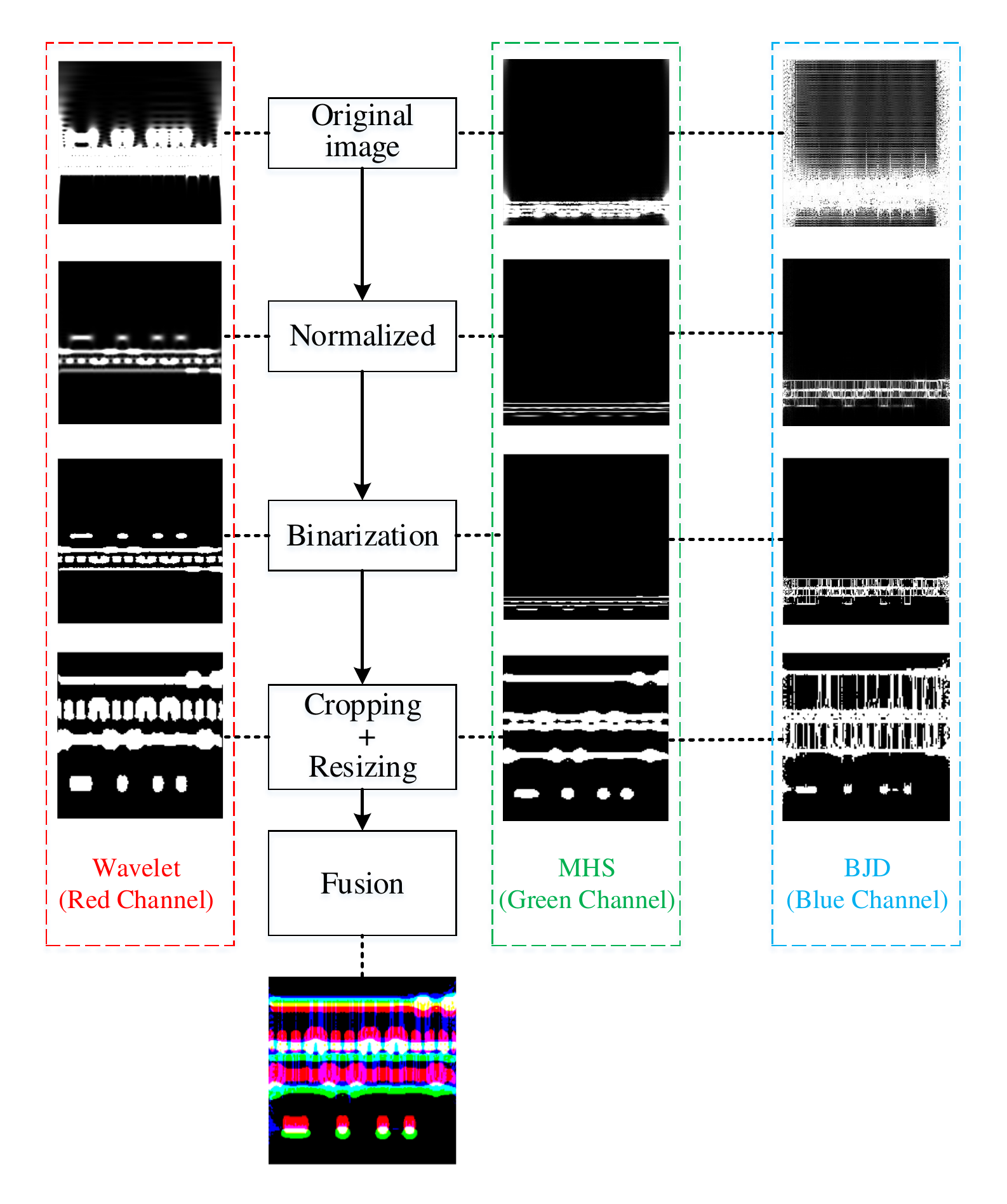}}
\captionsetup{font={footnotesize}}
\caption{$~$Flow chart of composite time-frequency analysis.}
\label{Fig.2.}
\end{figure}

Therefore, for the received signals, above linear and bilinear time-frequency analysis methods, i.e., the wavelet transform, MHD transform and BJD transform, are used to calculate the spectrograms to formulate the inputs for the three channels of the deep networks, as shown in Fig. \ref{Fig.2.}. That is, the grayscale images obtained from these three time-frequency transforms are used as the red, green and blue (RGB) sub-images, respectively. Subsequently, during the deep network training, these sub-images are input into the neural network as three channels of the deep network. Herein, the three channels of the composite spectrograms retain the characteristics of three time-frequency analysis methods, including both linear and nonlinear time-frequency analysis characteristics. However, in order to better extract the features and perform training and classification by the deep network, some preprocessing steps are required, i.e., the image normalized, binarized, cropped and resized, which will be explained as below.

\subsection{Image Normalization}
Due to the fact that, by performing time-frequency transform, the pixel value of the gray-scale spectrogram $I(x,y)$ for different kinds of interference signals may significantly different, and the data (image) with larger pixel value will have a greater impact on training, thereby destroying the balance of the data set. In order to reduce the imbalance of the data set, the gray-scale spectrogram is normalized before inputting into the deep network. That is, for the gray-scale spectrogram $I_1(x,y)$, the normalization method is

\begin{equation}
\begin{aligned}
{I_1}(x,y)=\begin{cases}
0,&I(x,y)\le{a_{\min}}\\
\frac{I(x,y)-a_{\min}}{a_{\max}-a_{\min}},&a_{\min}\textless I(x,y)\textless a_{\max}\\
1,&I(x,y)\ge a_{\max},\\
\end{cases}
\label{eq:7}
\end{aligned}
\end{equation}
where $a_{\min}$ and $a_{\max}$ are the minimum and maximum thresholds. That is, the pixel value smaller than $a_{\min}$ is set to 0, and the pixel value larger than $a_{\max}$ is set to 1. In this article, $a_{\min}$ and $a_{\max}$ are set to 0 and 137, respectively.

\subsection{Image Binarization}
After the normalization of the image, we should further perform the binarization before the step of image cropping and resizing. The binarization algorithm used in this paper is based on a global threshold \cite{7949002}. In particular, the binarization algorithm is summarized as \textbf{Algorithm} \ref{alg:1} shown below.

\begin{algorithm}
\caption{Binarization algorithm}
\label{alg:1}
    \begin{algorithmic}[1]
    \STATE $\mathbf{Input}$: Normalized time-frequency grayscale image $I_1(x,y)$;
    \REPEAT
        \STATE $\mathbf{Initialize}$: Threshold $T =({\mathop {\max}\limits_{x,y}(I_1(x,y))+\mathop {\min}\limits_{x,y}(I_1(x,y))})\slash 2$;
        \STATE Use $T$ to divide image $I_1(x,y)$ into two parts $H_1$ and $H_2$, where $H_1$ includes pixels with a value greater than $T$, and $H_2$ includes other values;
        \STATE Calculate the average values $\mu_1$ and $\mu_2$ of $H_1$ and $H_2$, respectively;
        \STATE Update $T=(\mu_1+\mu_2)\slash 2$;
    \UNTIL{$\Delta T\le 0.001$};
    \STATE $I_B(x,y)=\begin{cases}
    1,I_1(x,y)\ge T,\\
    0,otherwise.\\
    \end{cases}$;
    \STATE $\mathbf{Output}$: $I_B(x,y)$.
    \end{algorithmic}
\end{algorithm}
From the above, we can note that, the core of the algorithm is to calculate the binarization threshold $T$ and after the binarization, the value of each pixel is either 0 or 1.

\subsection{Image Cropping and Resizing}
From Fig. \ref{Fig.2.}, we note that most areas of the image after binarization contain no useful features. Therefore, in order to reduce redundant information of the image, part of the image has no useful information is cropped. That is, we carry out frequency domain clipping according to the start and end frequencies of the frequency hopping signal and also the highest and lowest frequencies of the interference signal, but does not cut in the time domain. In addition, in order to further reduce the amount of data and adapt to the input of the deep network, the size of the cropped image is adjusted so that all the images generated by the time-frequency transforms have the same size. Specifically, we use the nearest neighbor interpolation algorithm in image resizing as \cite{2008Digital}.

To sum up, for each sample signal (received by the frequency hopping receiver), at first, we perform the wavelet based linear time-frequency transform, MHD and BJD based bilinear time-frequency transforms to obtain three gray-scale spectrograms. Then aforementioned normalization, binarization, cropping and size adjustment are sequentially used for the three gray-scale spectrograms. In order to make full use of the characteristics of linear and Cohen-like bilinear time-frequency transforms, the three single-channel gray-scale time-frequency images are transferred to the red, green and blue colored images, respectively, to formulate the three-channels RGB color spectrograms and finally input into the deep neural network for classification.

\section{Siamese neural network}
The Siamese neural network was first introduced by Bromley and LeCun in the early 1990s to solve the problem of signature verification in the form of image matching \cite{1993Signature}. The Siamese neural network is composed by two sub-networks that accept different inputs and a top energy function. The energy function at the top calculates the distance between the topmost output feature representations of the sub-networks on both sides. The parameters between the two sub-networks are bound, so that the two sub-networks calculate the characteristics of their respective input samples according to the same rules \cite{koch2015siamese}. In the following, the architecture and the parameters of the Siamese neural network used in this paper are presented.
%
%
\subsection{Architecture of the Siamese Neural Network}
\begin{figure}[h]
\centerline{\includegraphics[scale=0.45]{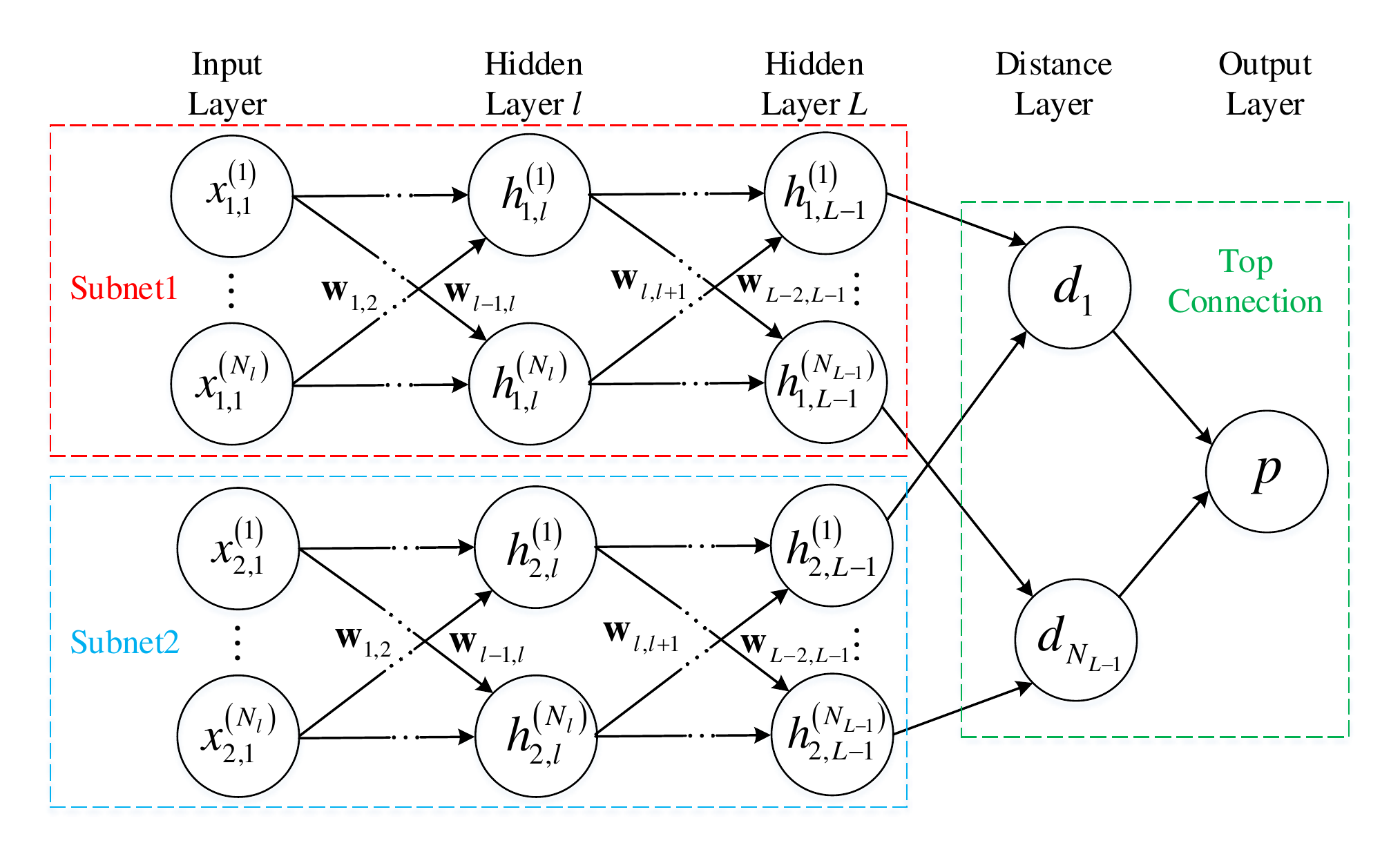}}
\captionsetup{font={footnotesize}}
\caption{$~$Architecture of the Siamese neural network.}
\label{Fig.3.}
\end{figure}

\begin{figure*}[]
\centerline{\includegraphics[scale=0.8]{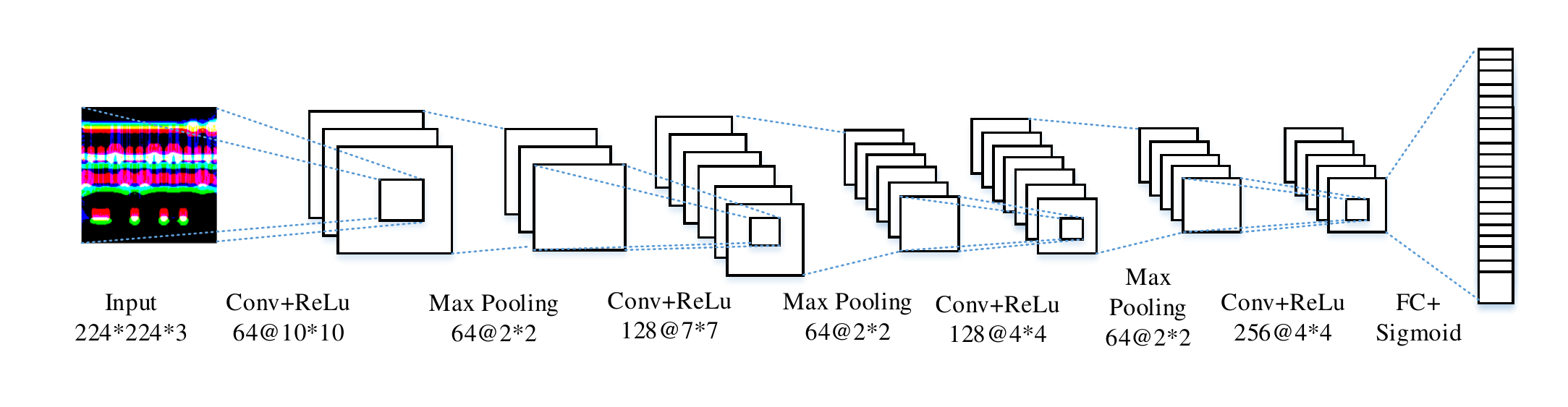}}
\captionsetup{font={footnotesize}}
\caption{$~$Architecture of the sub-network of the Siamese neural network.}
\label{Fig.4.}
\end{figure*}

The architecture of the Siamese neural network is shown in Fig. \ref{Fig.3.}. It contains $L$ layers, and each layer of the sub-network contains $N_l$ neurons, where $\mathbf{h}_{1,l}$ represents the hidden layer vector of the $l$th layer of sub-network $1$, and $\mathbf{h}_{2,l}$ represents the hidden hidden layer vector of the $l$th layer of sub-network $2$. The rectified linear (ReLU) unit is used as the activation function in the first $L-1$ layers of the Siamese neural network. Therefore, $\forall l\in\{1,\ldots,L-1\}$, we have the following relationship for the Siamese neural network,

\begin{equation}
h_{1,m}=\max(0,{\mathbf{W}_{l-1,l}^T}{\mathbf{h}_{1,(l-1)}}+{\mathbf{b}_l}),
\label{eq:8}
\end{equation}

\begin{equation}
h_{2,m}=\max(0,{\mathbf{W}_{l-1,l}^T}{\mathbf{h}_{2,(l-1)}}+{\mathbf{b}_l}).
\label{eq:9}
\end{equation}
${\mathbf{W}_{l-1,l}}$ represents the shared weight matrix of the two sub-networks connecting the $N_{l-1}$ neurons of the $l-1$th layer and the $N_l$ neurons of the $l$th layer, the size is ${N_{l-1}}\times{N_l}$, and ${\mathbf{b}_l}$ is the bias vector of the $l$th layer.

After the $L-1$ layers feed forward network, we then use the distance layer to quantify the difference of the eigenvectors $\mathbf{h}_{1,L-1}$ and $\mathbf{h}_{2,L-1}$ calculated by the two sub-networks. The distance function used by the distance layer is

\begin{equation}
p=\sigma\left(\sum\nolimits_j{\alpha _j}\left|{\mathbf{h}_{1,l}^{(j)}- {\mathbf{h}_{1,l}^{(j)}}}\right|\right),
\label{eq:10}
\end{equation}
where $\sigma(\cdot)$ is the sigmoidal activation function. This final distance layer induces a metric on the learned feature space of the $L-1$th hidden layer, and scores the similarity between the two feature vectors, and finally obtains the predicted score $p$. $\alpha_j$ are additional parameters learned by the model during the training process, which weighting the importance of the component-wise distance. This defines a final $L$th fully-connected layer for the network which joints the two Siamese twins.

\subsection{Parameters of the Siamese Neural Network}
The sub-network architecture of the Siamese neural network used in this paper is shown in Fig. \ref{Fig.4.}. The parameters of the two sub-networks are bounded and they have the same network architecture. That is, each sub-network is composed of 4 convolutional layers. Each convolutional layer uses a different size of convolution kernel, and the fixed step size is 1. The number of convolution kernels is designated as a multiple of 16, which can facilitate the training process and optimize the performance. The middle convolutional layer of the sub-network uses the ReLU function as the activation function to perform output feature mapping and then it is followed by a maximum pooling layer with a step size of 2. While after the last convolutional layer is a fully connected layer, the sigmoidal activation function is used for mapping for the fully connected layer. The Adam optimizer\cite{Adam} is adopted for the sub-network training, and the learning rate is dynamically adjusted. By gradually reducing the learning rate, the network can more easily converge to a minimum.

\section{Simulation Results}
In this section, the performance of the proposed compound time-frequency analysis and Siamese neural network based interference classification algorithm is evaluated and analyzed. Herein, we first introduce the training parameters for the Siamese neural network and the parameters used to generate the interference signal data set. Then we present some time-frequency analysis results of the interference signals under different scenarios and analyze how does the spectrograms are affected by the system parameters. Finally, the interference classification performance of the proposed algorithm is verified and analyzed by simulations.

\subsection{Simulation Parameters and Data Set Construction}
\subsubsection{Siamese Network Training Parameters}
The parameters used for training of Siamese neural network are listed in TABLE \,\ref{tab:1}. The parameters adopted by the two sub-networks are bounded and updated synchronously to perform fairly feature mapping for the input images. Herein, $N(\mu,\sigma^2)$ represents a normal distribution with mean value $\mu$ and variance $\sigma^2$.

\begin{table}[htbp]
\centering
\captionsetup{font={footnotesize}}
\caption{\,Training parameters of the Siamese network}
\begin{tabular}{cc}
  \hline
  \toprule Parameters & Value\\
  \midrule
  Weighting initialization value & $N(0,0.01)$ \\
  Bias initialization value      & $N(0.5,0.01)$ \\
  Learning rate initialization value &$6e-5$ \\
  \bottomrule
  \hline
\end{tabular}
\label{tab:1}
\end{table}

\subsubsection{Parameters for Signal Sample Generation}
The parameters used in the simulations for the frequency hopping communication system are shown in TABLE \,\ref{tab:2}, as \cite{20}, we consider a low-speed frequency hopping system.

\begin{table}[h]
\centering
\captionsetup{font={footnotesize}}
\caption{\,Parameters of frequency hopping signal}
\begin{tabular}{cc}
  \hline
  \toprule Parameters & Value\\
  \midrule
  Signal modulation method      & 2PSK \\
  Number of hopping frequency & 16 \\
  Frequency set                 &[100,220]kHz \\
  Rate                          & 100 hops per second \\
  Sampling frequency            &16MHz \\
  \bottomrule
  \hline
\end{tabular}
\label{tab:2}
\end{table}

While the parameters for the interference signals are set as TABLE \,\ref{tab:3} \cite{20}, in which $U[a, b]$ represents the uniform distribution on the interval $[a, b]$. Four kinds of interference are considered, i.e., fixed frequency interference, periodic linear sweep interference, periodic pulse interference and comb spectrum interference.

\begin{table}[h]
\renewcommand\arraystretch{1.5}
\centering
\captionsetup{font={footnotesize}}
\caption{\,Parameters of interference signal}
\begin{tabular}{c|cc}
  \hline
  Type&Parameters&Value\\
  \cline{1-3}
  Fixed frequency&Frequency set&{80,160,200}kHz\\
  \cline{2-3}
  interference&Sweep bandwidth&$U[50, 100]$kHz \\
  \cline{1-3}
  Periodic & Sweep period & $U[1e-6, 5e-6]$s \\
  \cline{2-3}
  linear sweep& Start frequency & $U[0, 100]$kHz \\
  \cline{2-3}
  interference & Pulse period & $U[3e-5, 8e-5]$s \\
  \cline{1-3}
  Periodic pulse& Pulse period & $U[3e-5, 8e-5]$s \\
  \cline{2-3}
  interference& Duty ratio   & $[0.2,0.5]$\\
  \cline{1-3}
  Comb spectrum & Number of comb teeth & $[4,8]$\\
  \cline{2-3}
  interference& Comb frequency       &[90,210]kHz\\
  \hline
\end{tabular}
\label{tab:3}
\end{table}

Also, in simulations, the interference signal strength is measured by the jamming-to-signal power ratio (JSR) defined as (\ref{eq:11}), where $V_{Jamming}$ and $V_{Signal}$ denote the average amplitudes of the interference signal and the frequency hopping signal, respectively.

\begin{equation}
{\rm{JSR =  - 20}}\lg \frac{{{V_{Jamming}}}}{{{V_{Signal}}}}
\label{eq:11}
\end{equation}
To generate signal sample and data set, the value of JSR varies from -10dB to 20dB with an interval of 5dB, i.e., we have 7 values of JSR. In addition, besides the singularity interference shown in TABLE \,\ref{tab:3}, we also consider the combination interference in the simulations. However, for the combination interference, we only discuss a combination of two kinds of interference. Therefore, there are $C_4^1+C_4^2=10$ types of interference. Unless otherwise stated, the training data set is generated as follows, for each value of JSR and the type of interference, 100 signal samples are randomly generated, i.e., we totaly have $7\times10\times100=7000$ signal samples. In addition, when generating the data set, the amplitudes of all interference signals are uniformly distributed over $[0.5,1.5]$. It should be noted that, the data input to the Siamese network is in the paired form. In particular, we randomly select two samples in the training set as the matching pairs. If the two samples of the matching pair have the same type, the label of the matching pair is $1$, otherwise the label of the matching pair is $0$. Furthermore, the Siamese neural network is iteratively trained 2000 times, and 180 matching pairs are randomly generated from the training set for each iterative training, so a total of 360,000 training matching pairs are included.

\subsection{Time-frequency Spectrogram for the Signals}
In this subsection, the time-frequency spectrograms of the composite interference signals are analyzed. To facilitate the analysis of the simulation results, herein, the number of hopping frequencies for the considered system is only set to 4. 
\subsubsection{The Time-Frequency Spectrogram for Different Time-frequency Methods}
\begin{figure}[h]
\centering
\begin{minipage}{12cm}
\subfigure[Wavelet]{\includegraphics[scale=0.4]{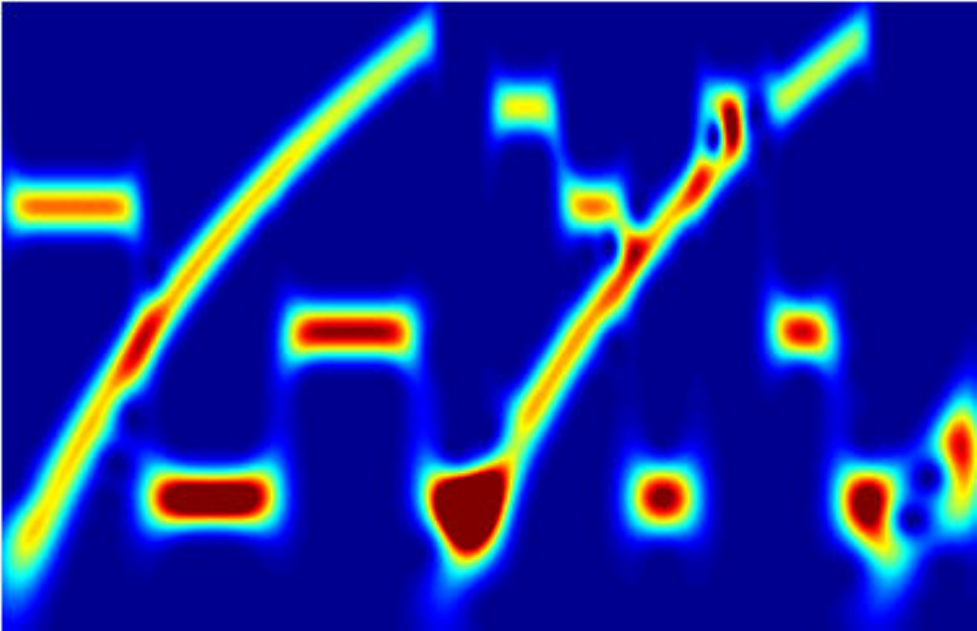}}
\subfigure[MHD]{\includegraphics[scale=0.4]{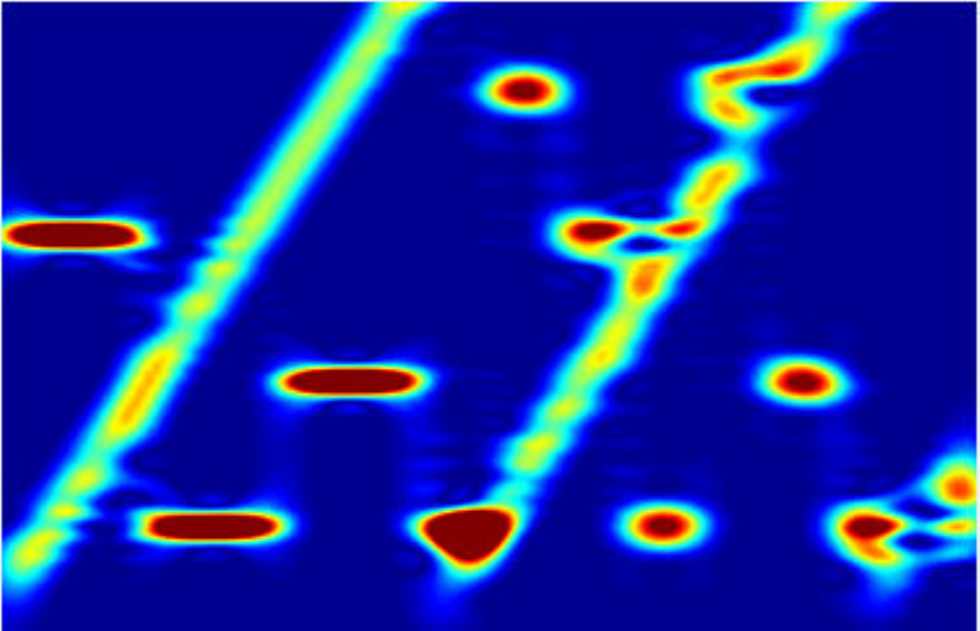}}
\end{minipage}
\begin{minipage}{12cm}
\subfigure[BJD]{\includegraphics[scale=0.4]{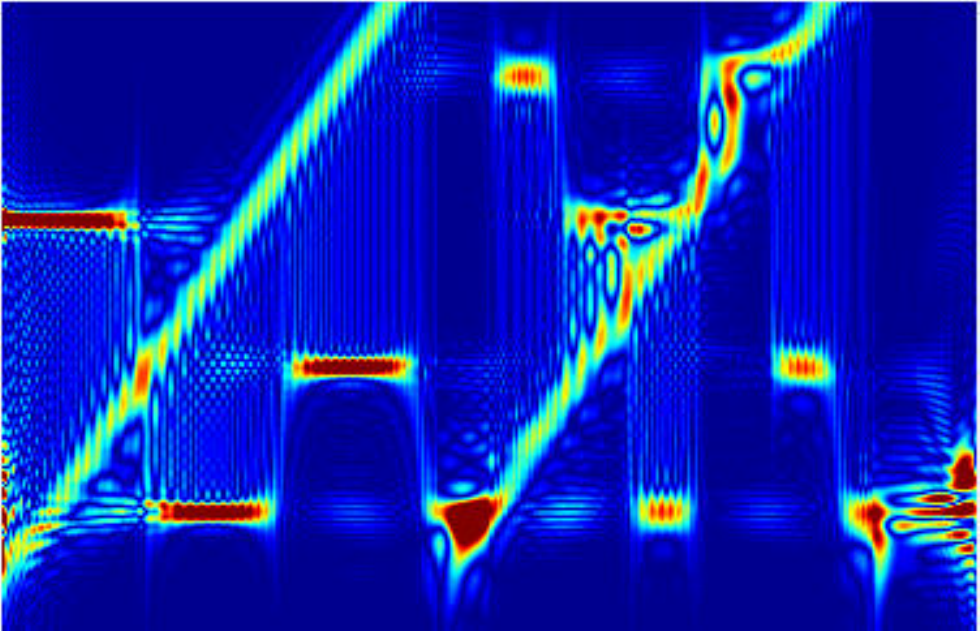}}
\subfigure[Wavelet, MHD and BJD]{\includegraphics[scale=0.4]{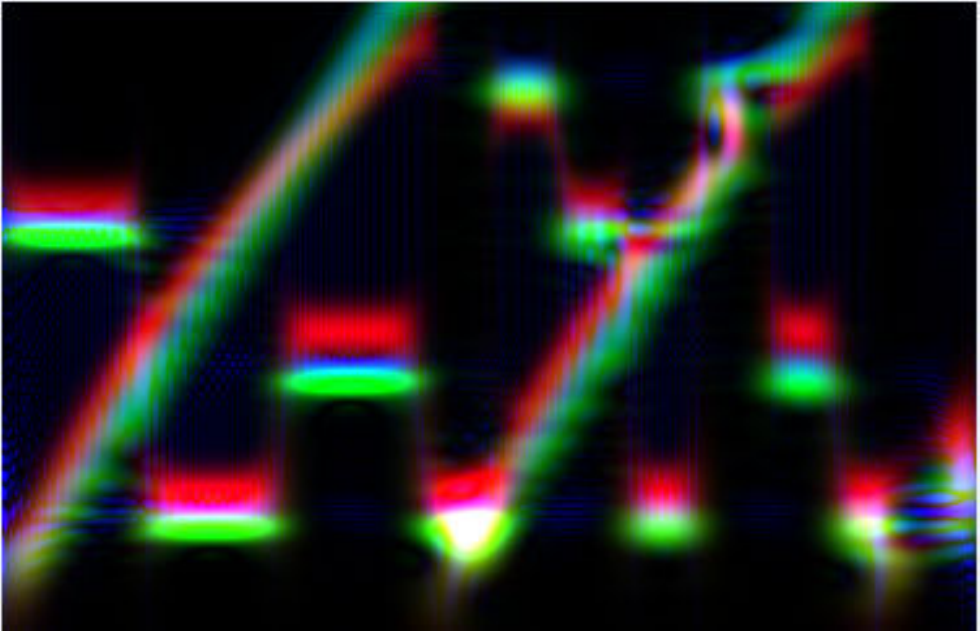}}
\end{minipage}

\captionsetup{font={footnotesize}}
\caption{$~$Spectrogram images of interference signals with different TFD.}
\label{Fig.5.}
\end{figure}

First, taking the frequency hopping signal interfered by the periodic linear sweep interference signal as an example, we analyze the time-frequency features extracted by different time-frequency analysis methods, i.e., Wavelet, MHD and BJD, in which the JSR is set to 0dB. The results are shown in Fig. \ref{Fig.5.}. While the horizontal axis of the time-frequency spectrogram is time and the vertical axis is frequency. The gradation of the color in the spectrogram indicates the normalized power intensity of the signal. It can be found from Fig. \ref{Fig.5.} that, for all the mentioned time-frequency analysis methods, the spectrograms of the frequency hopping signal exhibit the characteristics of time-varying and frequency hopping in the time-frequency domain, while the linear frequency sweeping signal shows the characteristic of linearly frequency changing with time in the time-frequency domain. Therefore, all the above-mentioned time-frequency analysis methods can extract the frequency hopping characteristics of the frequency hopping signal and the frequency gradual change characteristics of the periodic linear frequency sweep signal. Moreover, we can note that, the main difference among the spectrograms obtained by different time-frequency analysis methods is that when the signals overlap, the Cohen-like bilinear time-frequency representations, especially the BJD time-frequency transform using the kernel function $\phi({\eta,\tau})=\sin({\pi\tau\eta})\slash{(\pi \tau \eta)}$, has abundant cross terms. This is confirmed by the overlap part of the frequency hopping and frequency sweep signals in Fig. \ref{Fig.5.}(c). Compared with the features shown in Fig. \ref{Fig.5.}(a), BJD time-frequency analysis has more crossover features and fine image branches. In addition, for the proposed composite time-frequency analysis, the characteristics of linear and nonlinear time-frequency analysis are retained as that shown in Fig. \ref{Fig.5.}(d), which can be used to enhance the classification performance.

\subsubsection{Time-frequency Spectrogram for Different Kinds of Interference}
\begin{figure}[h]
\centering
\begin{minipage}{12cm}
\subfigure[Fixed frequency]{\includegraphics[scale=0.4]{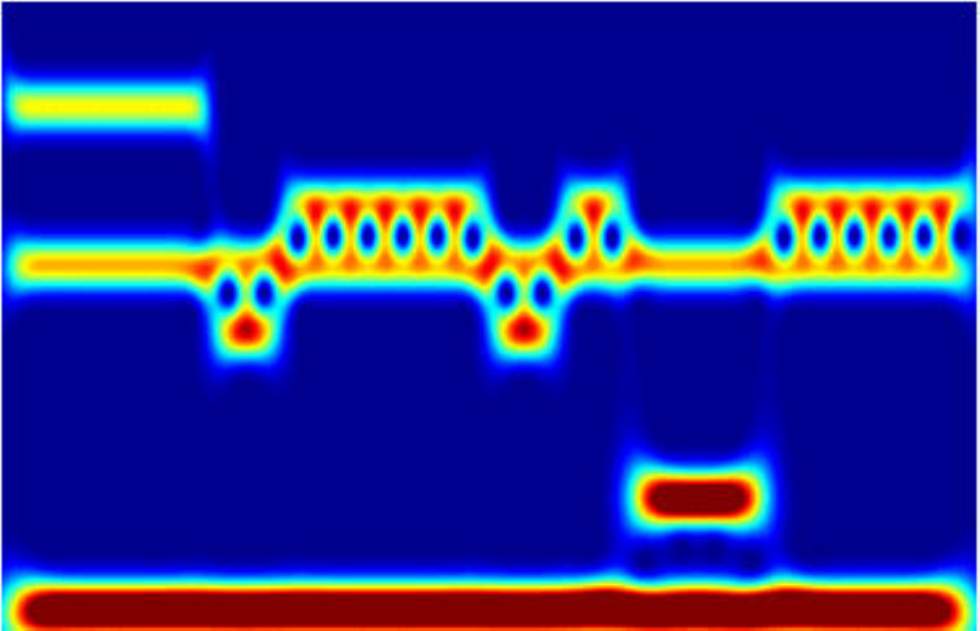}}
\subfigure[Periodic linear sweep]{\includegraphics[scale=0.4]{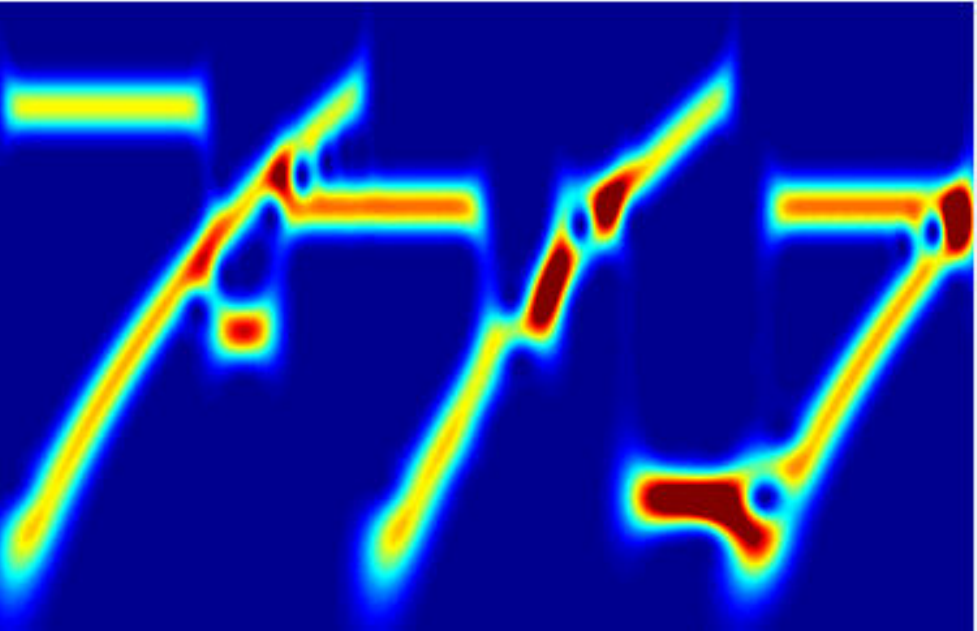}}
\end{minipage}
\begin{minipage}{12cm}
\subfigure[Periodic pulse]{\includegraphics[scale=0.4]{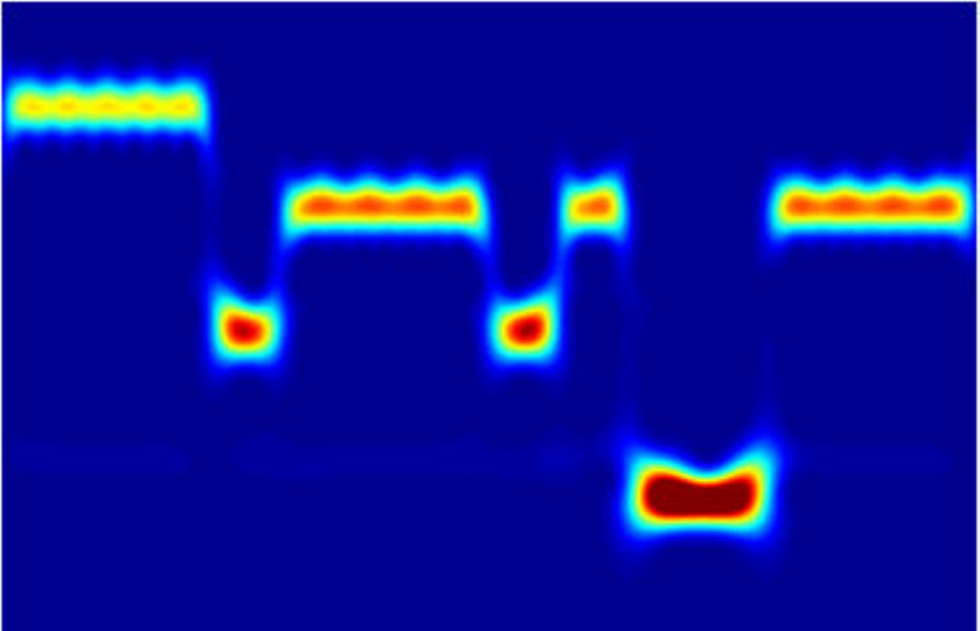}}
\subfigure[Comb spectrum]{\includegraphics[scale=0.4]{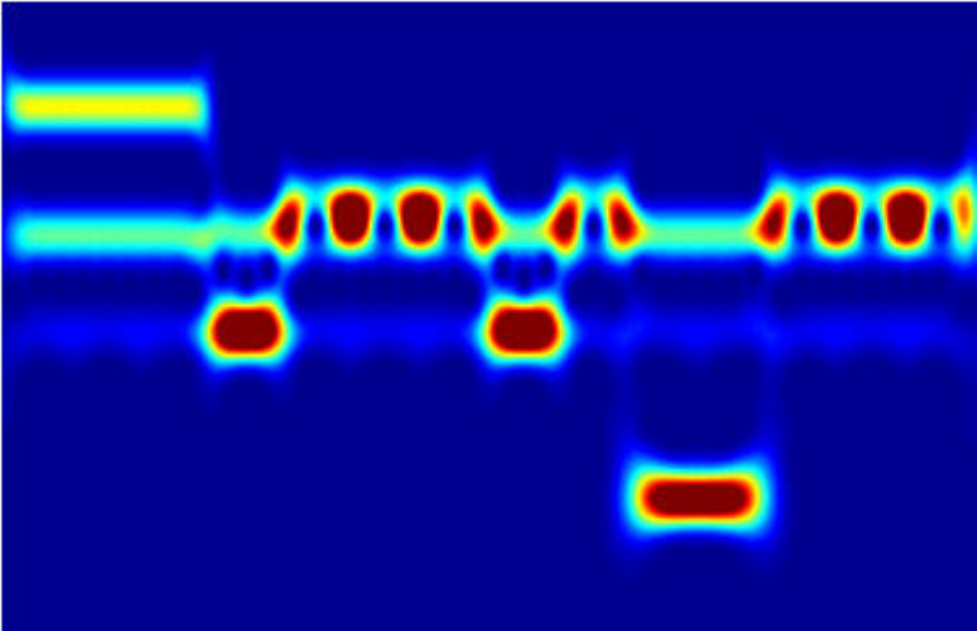}}
\end{minipage}
\begin{minipage}{12cm}
\subfigure[Fixed frequency and sweep]{\includegraphics[scale=0.4]{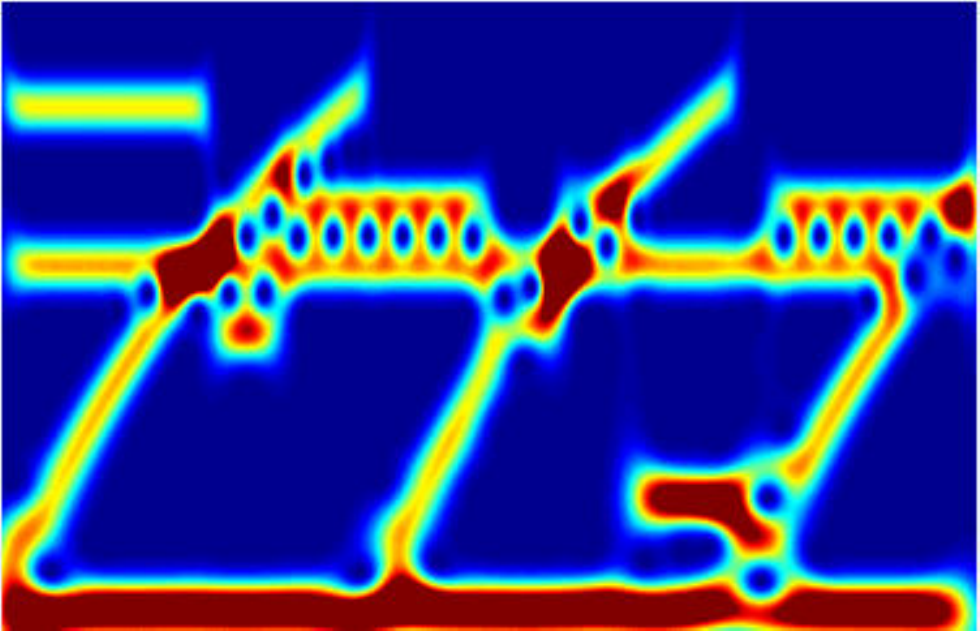}}
\subfigure[Pulse and comb spectrum]{\includegraphics[scale=0.4]{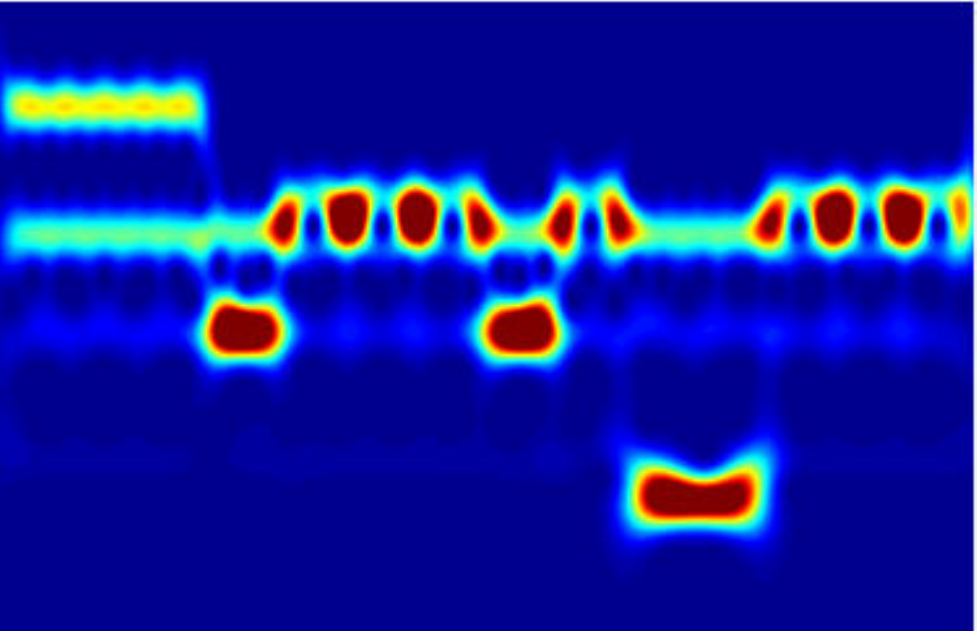}}
\end{minipage}

\captionsetup{font={footnotesize}}
\caption{$~$Spectrogram images of different interference signals.}
\label{Fig.6.}
\end{figure}

Now, the time-frequency spectrogram of the signals interfered by one of the aforementioned four kinds of singularity interferences or two kinds of composite interferences are analyzed. Herein, the JSR of the signals are set to 0dB and the Wavelet transform is taken as an example. The results are shown in Fig. \ref{Fig.6.}. In which, Fig. \ref{Fig.6.}(a) is the case that the signal is interfered by fixed-frequency interference. And from the time-frequency spectrogram, we can observe single or multiple interference signals on one fixed frequency or multiple frequencies, they are corresponding to single-tone and multi-tone interferences, respectively. We note that this type of interference has a great impacts for the frequency hopping signal appearing in its frequency range. For periodic linear frequency sweep interference, as shown in \ref{Fig.6.}(b), the interference signal is continuous in frequency domain and appears as a diagonal band in the spectrogram, which can interfere with a wide range of frequencies. The periodic pulse interference shown in Fig. \ref{Fig.6.}(c) contains multiple frequency bands with gradual power intensity in the spectrogram. The pulse period affects the distance between frequency bands in the spectrogram, and the duty cycle affects the distribution of distance between frequency bands in the spectrogram. For comb spectrum interference, as shown in Fig. \ref{Fig.6.}(d), in the interference frequency band, continuous interference frequency bands appear, corresponding to each comb tooth of comb spectrum interference.

In the case of multiple interferences coexisting, the time-frequency spectrogram will contain multiple interference characteristics. If the time-frequency characteristics of different types of interferences are significant different, it is still easy to perform interference classification, such as that shown in Fig. \ref{Fig.6.}(e), where the interference is composite by fixed frequency and linear frequency sweep signals. The problem becomes challenge as the composite interferences have similar time-frequency characteristics, such as that shown in Fig. \ref{Fig.6.}(f) where the interference is composite by periodic pulse and comb spectrum signals.

\subsubsection{The Influence of JSR}
\begin{figure}[h]
\centering
\begin{minipage}{12cm}
\subfigure[JSR=-10dB]{\includegraphics[scale=0.4]{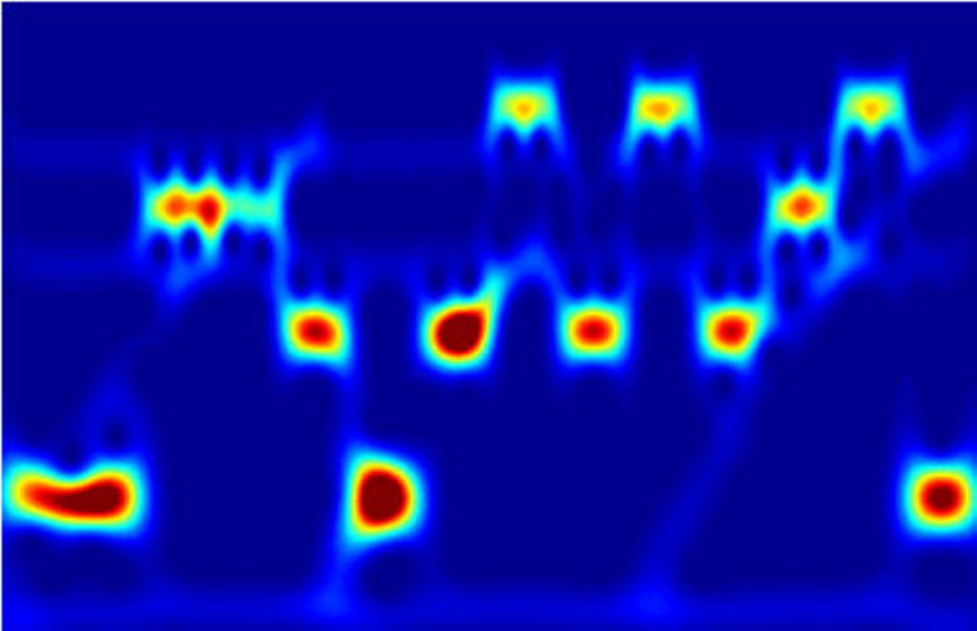}}
\subfigure[JSR=0dB]{\includegraphics[scale=0.4]{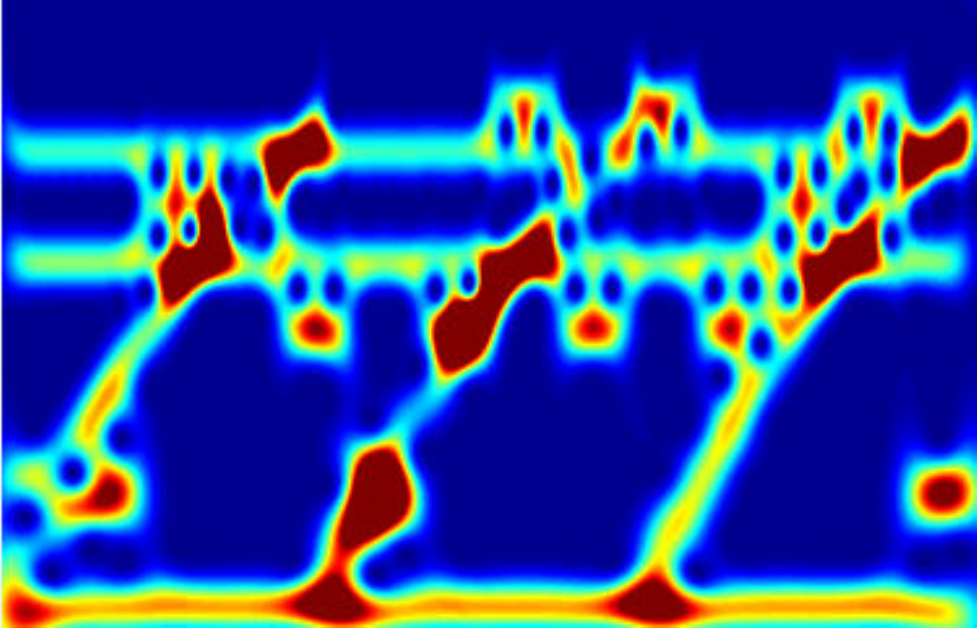}}
\end{minipage}
\begin{minipage}{12cm}
\subfigure[JSR=10dB]{\includegraphics[scale=0.4]{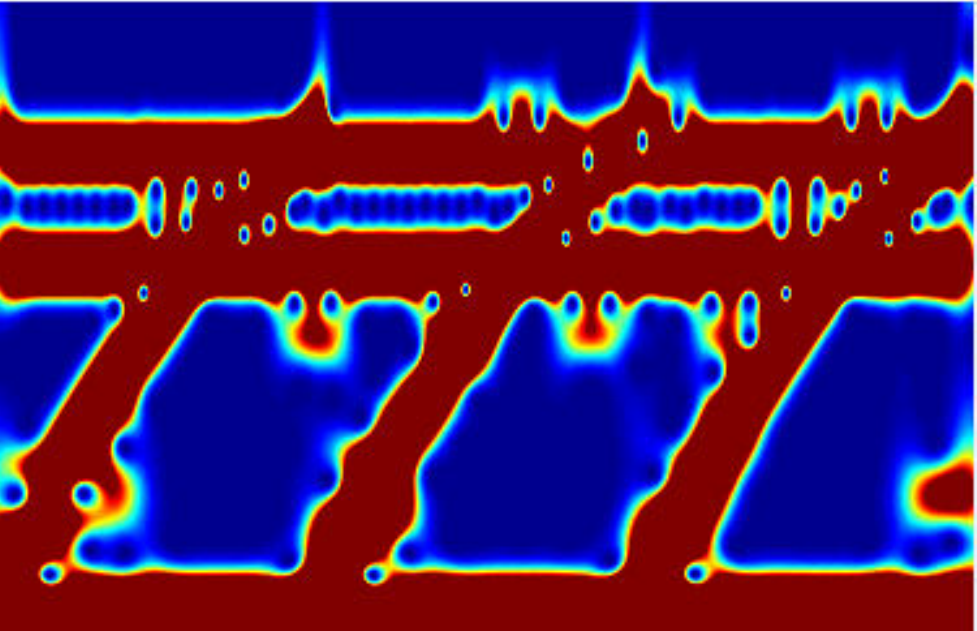}}
\subfigure[JSR=20dB]{\includegraphics[scale=0.4]{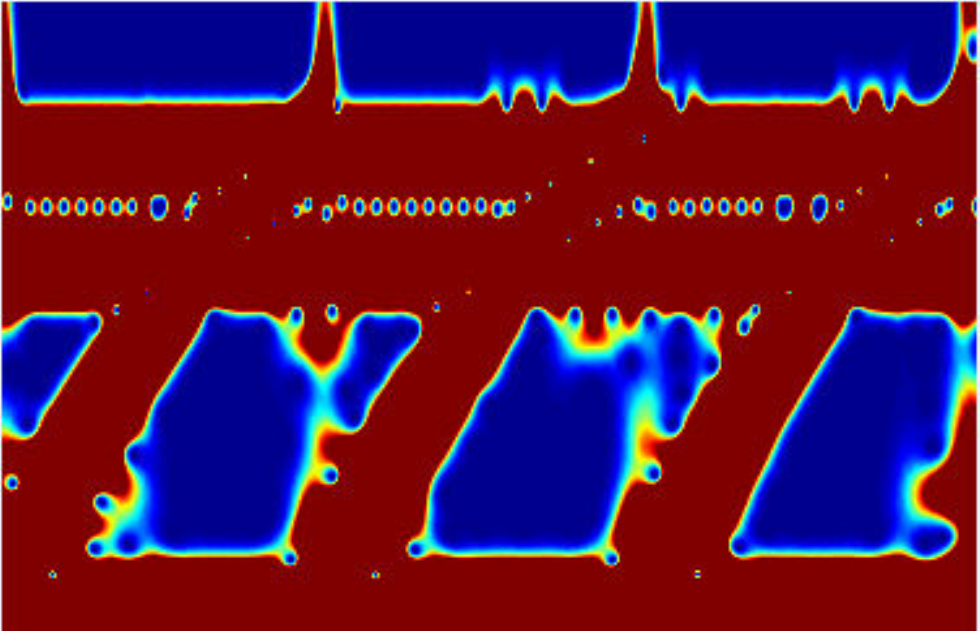}}
\end{minipage}

\captionsetup{font={footnotesize}}
\caption{$~$Spectrogram images of interference signals at different JSRs.}
\label{Fig.7.}
\end{figure}

At last, we analyze how does the time-frequency spectrogram is affected by interference signal intensity by varying the value of JSR and the result is shown in Fig.\ref{Fig.7.}. Herein, taking the frequency hopping system is interfered by the composite of sweeping and fixed frequency interferences as an example, and using Wavelet transform as the time-frequency analysis tool. It can be seen from the figure that with the increase of JSR, the color of the interference signal in the time-frequency spectrogram becomes darker and covers each other, especially as JSR=20dB, that is, the interferences are strong enough so that the features of different kinds of interferences are overlapped, thus the interference signal aliasing in the spectrogram and seriously reducing the distinguishability of them. While as JSR=-10dB, the interference intensity is low, now the interference is not obvious in the spectrogram and thus it is more difficult to distinguish the interference signals with similar characteristics. In fact, if JSR is small, there is no need to perform interference classification as it may not interfere the authorized communication system.

\subsection{Performance of the Algorithm}
In this subsection, the performance of the proposed interference classification algorithm is evaluated. At first, we explore how does the algorithm performance is affected by different time-frequency transforms. We then evaluate and analyze the algorithm performance by comparing it with traditional convolutional neural network and AlexNet transfer learning based interference identification algorithms. Finally, we further discuss the impact of sample data set size on the performance of the interference identification algorithm.

\subsubsection{Analysis the effect of composite time-frequency analysis method}
\begin{figure}[h]
\centerline{\includegraphics[scale=0.5]{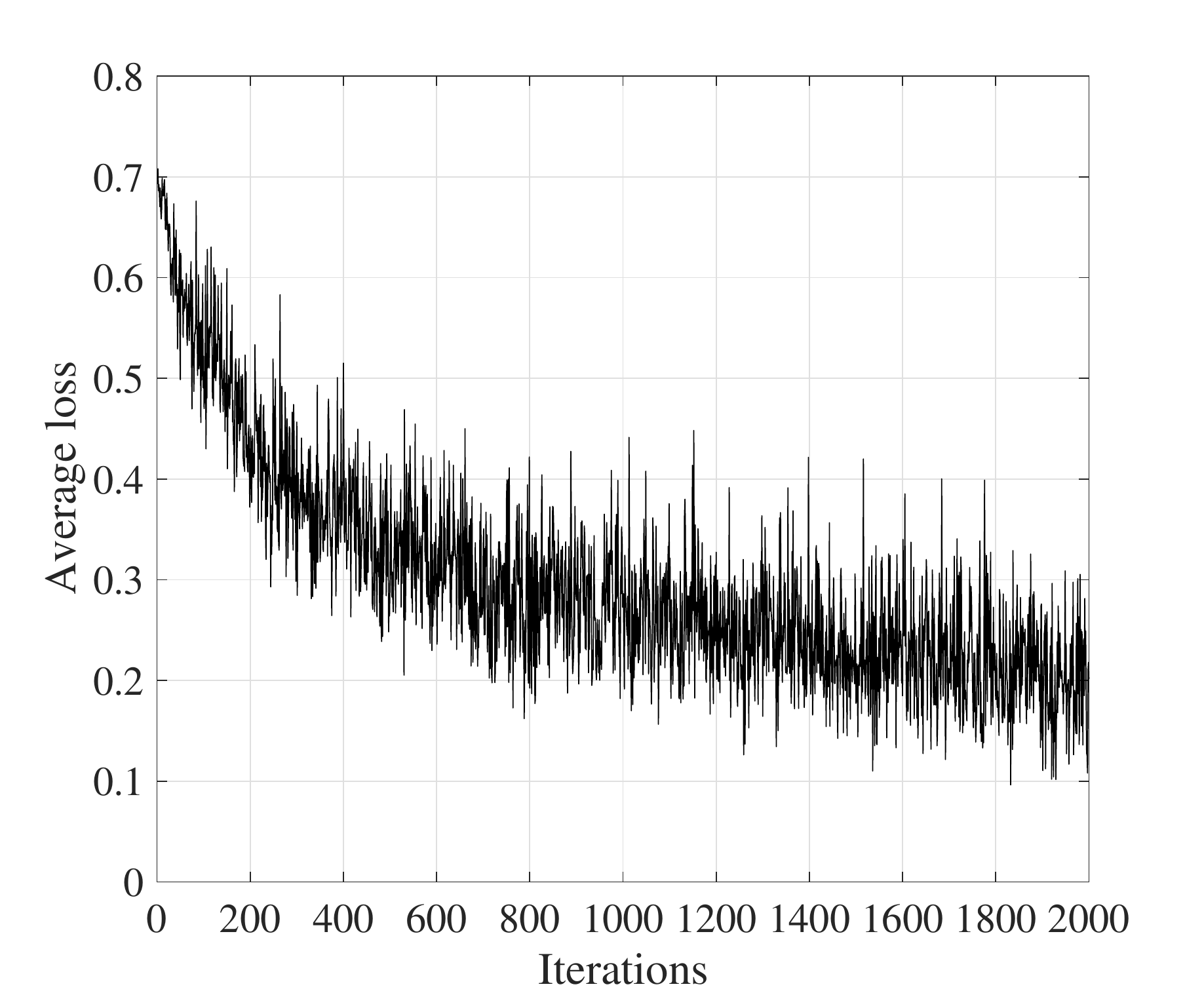}}
\captionsetup{font={footnotesize}}
\caption{$~$The training loss of the Siamese network.}
\label{Fig.8.}
\end{figure}

In order to verify the effect of the composite time-frequency analysis method on improving the performance of the interference classification algorithm, we analyze the classification accuracy of the algorithm based on linear, Cohen bilinear and composite time-frequency analysis method. The parameters used for training of the Siamese network and also the data set generation are as described in Section V-A. Fig. \ref{Fig.8.} shows the trend of the loss objective function during the training process of the Siamese network. It can be seen from the figure that after batch training, the loss function value gradually stabilizes, and the Siamese network gradually converges.

\begin{figure}[h]
\centerline{\includegraphics[scale=0.5]{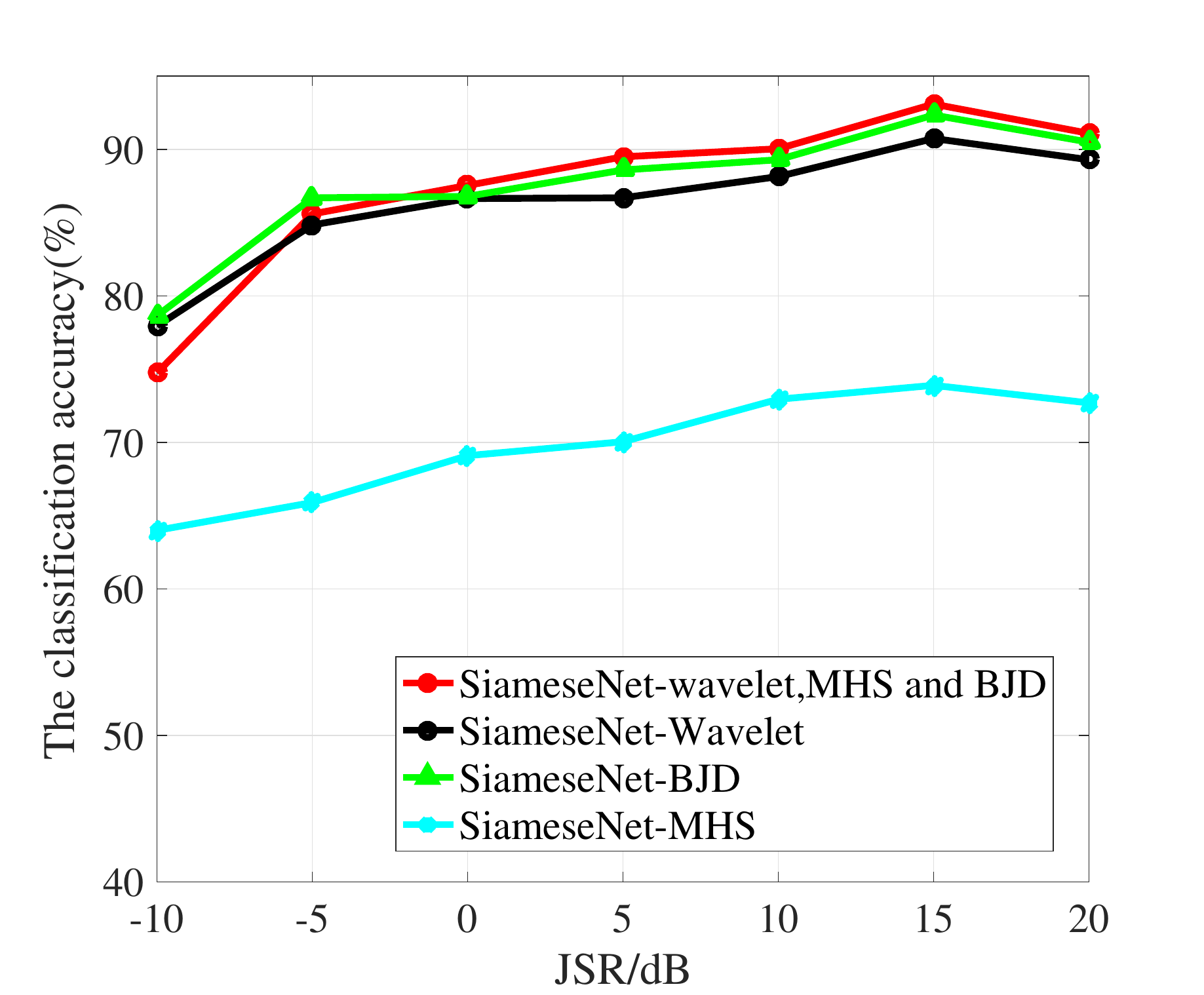}}
\captionsetup{font={footnotesize}}
\caption{$~$Classification accuracy of different time-frequency analysis methods versus JSR.}
\label{Fig.9.}
\end{figure}

Fig. \ref{Fig.9.} shows how the classification accuracy is affected by different time-frequency analysis methods, where the average classification accuracy is calculated herein and it is an average accuracy for identifying 10 kinds of interference. It can be seen from the figure that if JSR does not exceed 15dB, as JSR increases, the classification accuracy of different time-frequency analysis methods are all gradually improved, i.e., the intensity of interference increases and the features of the interference signals become obvious which facilitates the classification. While if JSR exceeds 15dB, the classification accuracy decreases. This phenomenon comes from the fact that, if JSR becomes larger, the interference signal intensity will be too larger so that a large area of the spectrogram is covered by the high-power signal features and the pixels color of the spectrogram becomes darker, i.e., part of the unique features of interference signal will be concealed, especially when the interferences are overlapped. This has been verified by time-frequency spectrogram shown in Section V-II.

From Fig. \ref{Fig.9.}, we can observe that, MHD based method has the lowest classification accuracy, followed by wavelet based method, and BJD based method obtains the highest accuracy among the single time-frequency analysis based methods. Also we note that, when the interference signal intensity is large, i.e., JSR$ \ge $0dB, the composite time-frequency analysis based interference classification algorithm obtains the highest classification accuracy. While the situation becomes worse as JSR$\textless $0dB, that is, as the interference signal intensity is too weak, the classification accuracy of the compound time-frequency analysis based method is slightly lower than the BJD time-frequency analysis based method. This result comes from the fact that, as the interference signal intensity is very low, the cross term generated by the BJD time-frequency transform can enhance the distinguishability of different kinds of interferences. However, though BJD, MHD and Wavelet transforms are combined in our composite time-frequency analysis method, the features obtained from this composite time-frequency analysis may be ignored thus achieve worse performance. In spite of this, this result still confirms that in most cases, the proposed composite time-frequency analysis based algorithm can obtain better performance than the other single time-frequency analysis based approaches.


\subsubsection{Analysis the effect of Siamese Network}
In order to fully characterize the performance of the proposed algorithm, we further analyze the effect of Siamese network and its performance by comparing it with the AlexNet transfer learning and the convolutional neural network based algorithms. The parameters of AlexNet can be found in \cite{2012ImageNet}. The core idea of the AlexNet transfer learning based interference classification algorithm is that we use the pre-trained AlexNet network and adaptive the last three layers of the network to our interference classification problem, and then the same data set are used to fine-tune the parameters of obtained AlexNet network. The configurations for the convolutional neural network are three convolutional layers with an $ReLU$ active function for each convolutional layer, then followed by a maximum pooling layer, and the last layer of the whole network is a fully connected layer. In comparison, the same data set described in Section V-A is used by the three networks for testing and training.

\begin{figure}[h]
\centerline{\includegraphics[scale=0.5]{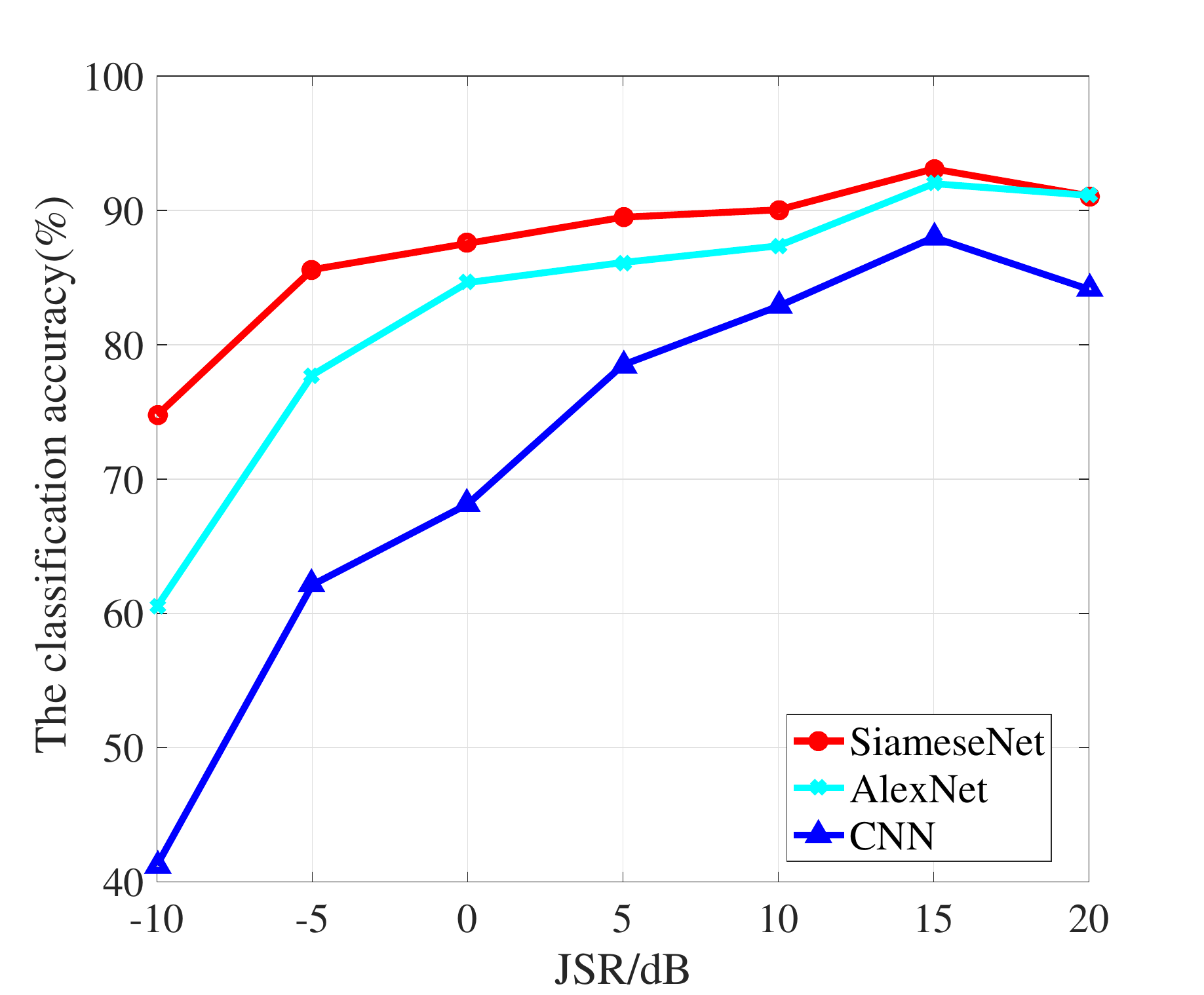}}
\captionsetup{font={footnotesize}}
\caption{$~$Classification accuracy of different neural network versus JSR.}
\label{Fig.10.}
\end{figure}

Fig. \ref{Fig.10.} shows the classification results of aforementioned three networks. Still, it can be seen from the figure that as JSR does not exceed 15dB, as JSR increases, the classification accuracy of all algorithms gradually increases. While as JSR exceeds 15dB, the classification accuracy of all three networks are reduced. This phenomenon is consistent with that shown in Fig. \ref{Fig.9.} and has the same reasons thus omitted for simplification. In addition, we also can note that the Siamese network based algorithm obtains the highest classification accuracy, followed by the AlexNet transfer learning based algorithm, and the convolutional neural network based algorithm has the lowest classification accuracy. This result verifies the dominated performance of the proposed Siamese network based algorithm over the AlexNet transfer learning and the convolutional neural network based algorithms.

\subsubsection{Analysis the impact of sample data set size}
In order to evaluate the performance of the algorithm on small data sets, we further demonstrate the classification performance of the Siamese network with different training matching pairs. Under the condition that the total number of training samples is unchanged, TABLE \,\ref{tab:4} shows how does the classification accuracy is varying with the training matching pairs of the Siamese network. The average classification accuracy in the table is an average interference classification accuracy under various JSR. One can note that, as the number of matching pairs decreases, the average classification accuracy decreases, while as the number of iterations and the number of matching pairs in each batch of training decreases, the classification accuracy also decreases. However, the reduction in the number of matching pairs does not significantly reduce the classification accuracy. In addition, it is worth mentioning that the average classification accuracy of the aforementioned two comparison algorithms, namely, the AlexNet transfer learning and the convolutional neural network based algorithms are 82.79\% and 72.14\%, respectively. It can be seen that even if fewer matching pairs are used by the Siamese network during the training process, its interference classification accuracy is still higher than the AlexNet transfer learning and the convolutional neural network based algorithms. The main reason is that, in order to generate training matching pair samples, the random matching is used by the Siamese network, thus the trained Siamese network is more robust in interference classification than the AlexNet transfer learning and the convolutional neural network based algorithms. This result also shows the dominated performance of the proposed Siamese network based algorithm.

\begin{table}[htbp]
\centering
\captionsetup{font={footnotesize}}
\caption{\,Effects of the number of training sample and matching pair on Siamese network}
\begin{tabular}{p{1cm}cp{1cm}p{1.2cm}p{1.5cm}}
  \hline
  \toprule Training samples & Iterations & Matching pairs per iteration & Total matching pairs&Average recognition accuracy(\%)\\
  \midrule
  7000 & 2000 & 180 & 360000 & 87.38\\
  7000 & 2000 & 100 & 200000 & 86.99\\
  7000 & 1000 & 180 & 180000 & 86.34\\
  7000 & 1000 & 100 & 100000 & 85.92\\
  \bottomrule
  \hline
\end{tabular}
\label{tab:4}
\end{table}

\section{Conclusion}
In this paper, the interference classification problem for the frequency hopping communication systems are discussed and then a composite time-frequency analysis and Siamese neural network based interference classification algorithm is proposed. Specifically, in order to fully extract the characteristics of various interferences, at first, a composite time-frequency analysis algorithm has been designed to calculate the time-frequency distribution of the interference signal. Both the linear time-frequency transform, i.e., Wavelet transform, and the bilinear time-frequency transform, i.e., MHD and BJD transforms, are used to extract the time-frequency representations of the interference signals, and thus three-channel time-frequency spectrograms are obtained. Then before the multi-channel time-frequency spectrograms are input to the deep network for classification, they are normalized, binarized, resized and cropped. Finally, the Siamese network is selected as the classifier. In which, the twin sub-network of the Siamese neural network calculates two distance vectors of the input samples on each sub-network to determine whether the two inputs are the same type of samples. The matching and classification of the samples are realized after repeated training. From the simulation results, we found that the classification accuracy of the proposed algorithm based on composite time-frequency analysis is better than the method based on single time-frequency transformation in most cases. In addition, through comparison, we also have noted that the proposed algorithm obtains higher classification accuracy than both the AlexNet transfer learning and the convolutional neural network based methods.

\appendices
\section{Typical interference patterns}
According to the interference pattern, there are four kinds of interferences for the wireless communication system, i.e., fixed interference, sweep interference, periodic pulse interference and comb spectrum interference. The forms of these interference signals are presented below.

\emph{Fixed interference}: It performs continuous interference at a specific frequency, the interference signal form is
\begin{equation}
J(t)=\sum_{i=1}^{N} A_i\cos(2\pi f_it+\phi_i),
\label{eq: A.1}
\end{equation}
where $N$ denotes the number of interference frequencies, if $N=1$, it is degenerated to the monophonic interference, while if $N>1$, it is termed as multi-tone interference. In addition, $A_i$, $f_i$ and $\phi_i$ represent the amplitude, frequency and initial phase of the $i$th interference signal, respectively. The key parameters for the fixed frequency interference are $f_i$, $N$ and $A_i$.

\emph{Sweep interference}: It is a suppressive interference that scans each frequency in a designated frequency band to destroy all the authorized signals over this frequency band. In this paper, we only consider the periodic linear sweep interference and the signal form with single period is
\begin{equation}
J(t)= A\cos(2 \pi f_0 t + \pi\mu_0 t^2 + \phi_0).
\label{eq: A.2}
\end{equation}
Where $A$, $f_0$, $\mu_0$ and $\phi_0$ denote the amplitude, initial frequency, sweep slope and initial phase of the sweep interference, respectively. In practical, the sweep interference is periodically performed over a designated frequency band and the main parameters are the scanning cycle and sweep slope.

\emph{Periodic pulse interference}: It is an interference signal composed of a narrow pulse sequence. The signal form is
\begin{equation}
J(t)= \sum_{i} Ag_{\tau}(t-iT),
\label{eq: A.2}
\end{equation}
where $A$ and $T$ denote the amplitude and pulse repetition period of the periodic pulse interference signal, respectively, and $g_{\tau}(.)$ represents a rectangular pulse with pulse width $\tau$. In addition, for the periodic pulse interference, there is an inexplicit parameter, i.e., the duty factor $\gamma$, and it is defined as $\gamma=\tau/T$. The key parameters for the periodic pulse interference signals are the pulse width $\tau$, repetition period $T$ and the duty factor $\gamma$.

\emph{Comb spectrum interference}: It is a kind of discrete blocking interference and it has multiple discrete narrow band interferences over a designated frequency band. The signal form is
\begin{equation}
J(t)= \sum_{i=1}^{N} A_i(t)\cos(2\pi f_i t+\phi_i(t)),
\label{eq: A.2}
\end{equation}
where $N$, $A_i(t)$, $f_i$ and $\phi_i(t)$ represent the number of comb, envelope, comb frequency and phase of the comb frequency interference signals, respectively. We can note that, the comb frequency interference is a superposition of $N$ narrow band interference signals. For the comb interference signals, the bandwidth, envelope, the gap among the combs are all adjustable parameters.

\ifCLASSOPTIONcaptionsoff
  \newpage
\fi



\end{document}